\documentclass[iop,apjl]{emulateapj} \usepackage{times}

\newcommand{\gray}{$\gamma$-ray}
\newcommand{\grays}{$\gamma$-rays}

\newcommand{\adv}{Adv.\ Space Res.}

\newcommand{\arnps}{Ann.\ Rev.\ Nuc.\ Part.\ Sci.\ } 
\newcommand{\pubjournal}[5]{#4, #1, #2, #3}

\newcommand{\fermi}{{\it Fermi}}
\newcommand{\fermilat}{{\it Fermi}--LAT}
\shorttitle{Luminosity and energy budget of the Milky Way}
\shortauthors{Strong et al.}

\begin{document}

\title{Global cosmic-ray related luminosity and energy budget of the Milky Way}

\author{A.~W. Strong\altaffilmark{1}}
\altaffiltext{1}
%\affil
{Max-Planck-Institut f\"ur extraterrestrische Physik,
Postfach 1312, D-85741 Garching, Germany
\email{aws@mpe.mpg.de}}

\author{T.~A. Porter\altaffilmark{2}}
\altaffiltext{2}
%\affil
{Hansen Experimental Physics Laboratory, 
  Stanford University, Stanford, CA 94305
   \email{tporter@stanford.edu}}

\author{S.~W. Digel\altaffilmark{3,4}}
\altaffiltext{3}
%\affil
 {SLAC National Accelerator Laboratory, 
  2575 Sand Hill Road, Menlo Park, CA 94025
   \email{digel@slac.stanford.edu}}
\altaffiltext{4}
{Kavli Institute for Particle Astrophysics and Cosmology,
Stanford University, Stanford, CA 94305}

\author{G. J\'{o}hannesson\altaffilmark{2}} %\email{gudlaugu@stanford.edu}

\author{P. Martin\altaffilmark{1}} %\email{martinp@mpe.mpg.de}

\author{I.~V. Moskalenko\altaffilmark{2,4}} %\email{imos@stanford.edu}

\author{E.~J. Murphy\altaffilmark{5}}
\altaffiltext{5}
{Spitzer Science Center,
  California Institute of Technology, Pasadena, CA 91125 
\email{emurphy@ipac.caltech.edu}}

%\and

\author{E. Orlando\altaffilmark{1}} % \email{elena.orlando@mpe.mpg.de}

\begin{abstract}
We use the GALPROP code for cosmic-ray (CR) propagation to
calculate the broad-band luminosity spectrum
of the Milky Way related to CR propagation and interactions 
in the interstellar medium. 
This includes \gray{} emission from the production and subsequent 
decay of neutral pions ($\pi^0$), bremsstrahlung, and 
inverse Compton scattering, and synchrotron radiation. 
The Galaxy is found to be nearly a CR electron calorimeter, 
but {\it only} if \gray{} emitting processes are taken into account. 
Synchrotron radiation alone accounts for only one third of the total 
electron energy losses with $\sim 10-20$\% of the total synchrotron 
emission from secondary CR electrons and positrons. 
The relationship between far-infrared and radio luminosity that we find from 
our models is consistent with that found for galaxies in general.
The results will be useful for understanding the connection between diffuse 
emissions from radio through \gray{s} in ``normal'' (non-AGN dominated) 
galaxies, as well as for estimating the broad-band extragalactic diffuse 
background from these kinds of galaxies. 
\end{abstract}


\section{Introduction}
%##############################################################################

Cosmic rays (CRs) fill up the entire volume of galaxies, providing an important 
source of heating and ionisation of the interstellar medium (ISM), and may play 
a significant role in the regulation of star formation and evolution of 
galaxies \citep{Ferriere2001, Cox2005, Socrates2008, Sironi2010}.
Diffuse emissions from radio to high-energy \grays{} ($>$100 MeV)
arising from various interactions between CRs and the ISM, 
interstellar radiation field (ISRF), and magnetic field, are currently 
the best way to trace the intensities and spectra of CRs 
in the Milky Way (MW) and other galaxies. 
Gamma rays are particularly useful in this respect since this energy 
range gives access to the dominant hadronic component
in CRs via the observation of $\pi^0$-decay radiation produced by 
CR nuclei inelastically colliding with the interstellar gas.
Understanding the global energy budget of processes related to the injection
and propagation of CRs, and how the energy is distributed across the 
electromagnetic spectrum, is essential to interpret the 
radio/far-infrared relation \citep{deJong1985,Helou1985,Murphy2006}, galactic 
calorimetry \citep[e.g.,][]{Volk1989}, and predictions of 
extragalactic backgrounds \citep[e.g.,][]{Thompson2007,Murphy2008}, and for
many other studies.

The MW is the best studied non-AGN dominated star-forming 
galaxy, and the only galaxy that direct measurements of CR intensities and 
spectra are available.
However, because of our position inside, the derivation of 
global properties is not straightforward and requires detailed 
models of the spatial distribution of the emission.
Nevertheless, constructing a model for the global properties of such a galaxy
is tractable with the variety of data available.

In this paper, we calculate the injected CR power and corresponding 
broad-band luminosity spectrum from radio to \gray{s} for CR propagation
models consistent with current CR, radio, and \gray{} data.
Earlier estimates focussing on the MW \gray{} luminosity only 
\citep[e.g.,][]{SMR2000} were based on modelling EGRET data.
The launch of the Large Area Telescope (LAT) on the \fermi\ 
{\it Gamma-ray Space Telescope} (hereafter, \fermilat), has provided 
a wealth of 
new \gray{} data up to, and beyond, 100 GeV. 
Analysis of the \fermilat\ data has 
not confirmed the anomalous ``EGRET GeV-excess'' \citep{Abdo2009a}
and has led to an improved model of CR propagation and diffuse \gray{} emission
thus enabling a better estimate of the CR and \gray{} luminosities 
of the Galaxy.

\section{The Models}
%##############################################################################

We use a model based on the GALPROP 
code\footnote{http://galprop.stanford.edu} for the CR-related processes in 
the Galaxy that has been 
adjusted to fit many types of data including direct measurements of CRs, 
non-thermal radio emission, hard X-rays, and \gray{s}.
Studies of the diffuse Galactic emission and CRs using this code prior to the
launch of the \fermilat\ can be found in 
\cite{Moskalenko1998}, \cite{Strong1998}, \cite{SMR2000}, 
\cite{Moskalenko2002}, \cite{Strong2004a}, \cite{Strong2004b}, 
and \cite{Porter2008}.
An extensive review of CR propagation, models, data, and 
literature is given by \cite{Strong2007}. 

We calculate diffuse emissions from radio to \gray{s}, produced by CR protons, 
helium, and electrons/positrons.
The contributions by discrete sources and line 
emissions (e.g., 511-keV annihilation radiation) are 
not included.
Since the propagation parameters are not uniquely determined from the 
observations we consider 
an illustrative range of parameters for 
diffusive-reacceleration (DR) and plain diffusion (PD) propagation models.
The main uncertainty we consider is the CR confinement 
volume and associated 
propagation model parameters, e.g., the diffusion coefficient.
Other effects to consider would include different 
propagation modes, such as convection, 
the distribution of CR sources, etc. 
However, these are outside the scope of the current paper and we defer these
to future work.

The output of a typical GALPROP run includes
CR distributions, \gray{}/synchrotron skymaps and emissivity distributions. 
The luminosities of the different components are computed by integrating 
the respective emissivities over the total Galactic volume: for
IC and synchrotron radiation the volume emissivity is directly
calculated, while for $\pi^0$-decay
and bremsstrahlung the emissivity is calculated 
per hydrogen atom so we weight the 
calculated volume emissivities by the distributions of atomic, molecular, 
and ionised hydrogen included in the GALPROP code 
\citep{Strong1998,Moskalenko2002,Strong2004b}.

We obtain the {\it input} CR luminosity for each of our models by integrating
the model CR source spectra and spatial distribution over the Galaxy.
The injected luminosity, instead of emergent, is the relevant 
quantity because we are interested in how the CR power is transferred to the 
electromagnetic channels. 
In addition to CR protons and helium, and primary electrons, we also calculate 
the luminosities of secondary positrons and electrons resulting from CR-gas
interactions in the ISM.
The diffuse emissions from these secondary CR species are also calculated and 
included in the total synchrotron, 
bremsstrahlung, and IC spectra.

We use DR and PD propagation models 
where the spatial diffusion coefficient D$_{xx}$ and its momentum dependence, 
together with the diffusive reaccelation characterised by 
an Alfv\'en speed, $v_A$, assuming a Kolmogorov spectrum
of interstellar turbulence (if used), and the 
size of the CR confinement volume, $z_h$, 
are obtained by fitting the CR secondary/primary ratios for B/C 
and $^{10}$Be/$^9$Be; for details see \cite{Strong2007}.
The CR source injection spectra are taken as broken power-laws in 
momentum, with different parameters for nuclei and primary electrons. 
These are chosen to
reproduce the directly observed CR spectra after propagation for the 
adopted models. 

The calculations are based on model parameters that
reproduce the \fermilat\ electron spectrum \citep{Abdo2009a} 
and \gray{} data \citep{Abdo2009a,Abdo2009b,Abdo2010a,Abdo2010b}.
The spatial distribution of the CR sources is based on 
pulsars as tracers of 
supernova remnants (SNRs) 
used in \cite{Strong2004b}, but constant beyond the solar circle as 
indicated by analysis of the \fermilat\ data for the 
$2^{\rm nd}$ Galactic quadrant \citep{Abdo2010a}.

The synchrotron calculation uses the same CR electron model and a 
magnetic field model that reproduces radio surveys at frequencies from 
100 MHz to 23 GHz (Strong et al. {\it in preparation}). 
Diffusive-reacceleration and PD propagation models are used for 
three halo sizes, $z_h =$ 2, 4, and 10 kpc, respectively, with 
corresponding self-consistently derived diffusion 
coefficients $D_{xx}(\rho)=D_0 \beta (\rho/\rho_0)^\delta$, 
where $\beta=v/c$ is the dimensionless particle velocity, $\rho$ is the 
particle rigidity, and $D_0$, $\rho_0$, $\delta$ are constants.
These halo sizes cover the range consistent
with available CR data for B/C and $^{10}$Be/$^9$Be 
\citep[][and references therein]{Engelmann1990,Yanasak2001,deNolfo2006}.
The parameters and values for 
the models are summarised in Table~\ref{table:1}.

The optical to far infrared (FIR) 
luminosity spectrum is derived from the model of the 
ISRF used for the propagation calculations 
\citep[the ``maximum metallicity gradient'' model from][]{Porter2008}.
The emergent luminosity for the ISRF is computed by surface integration 
over a region large enough to encompass the total flux from the stellar 
luminosity distribution and the starlight reprocessed by dust 
($\sim$30 kpc radius about the Galactic centre).
In the present work, the input bolometric stellar luminosity is 
$\sim$$4 \times 10^{10} L_\sun$
apportioned across the stellar components 
boxy bulge/thin disc/thick disc/halo with
fractions $\sim$0.1/0.7/0.1/0.1, and $\sim$20\% reprocessed by dust and 
emitted in the infrared.
The uncertainties related to the 
distribution of the ISRF interior to the integration boundary include the 
relative luminosities of the bulge component and the disk, 
the metallicity gradient, and other details.
However, these uncertainties are dominated by the overall uncertainty in
the input stellar luminosity.
A higher input stellar luminosity will increase the CR electron/positron 
losses via IC scattering and hence the overall output in \gray{s}, 
requiring a higher injected CR power and increasing the overall calorimetric
efficiency (see below).
Estimates available in the literature illustrating the range for the 
MW stellar luminosity are, e.g., 
$6.7\times10^{10}$ $L_\odot$ \citep{Kent1991} and
$2.3\times10^{10}$ $L_\odot$ \citep{Freudenreich1998}.

%#############################################################################
\section{Results and Discussion}
%##############################################################################

\begin{figure}
\centerline{
\includegraphics[width=3.75in]{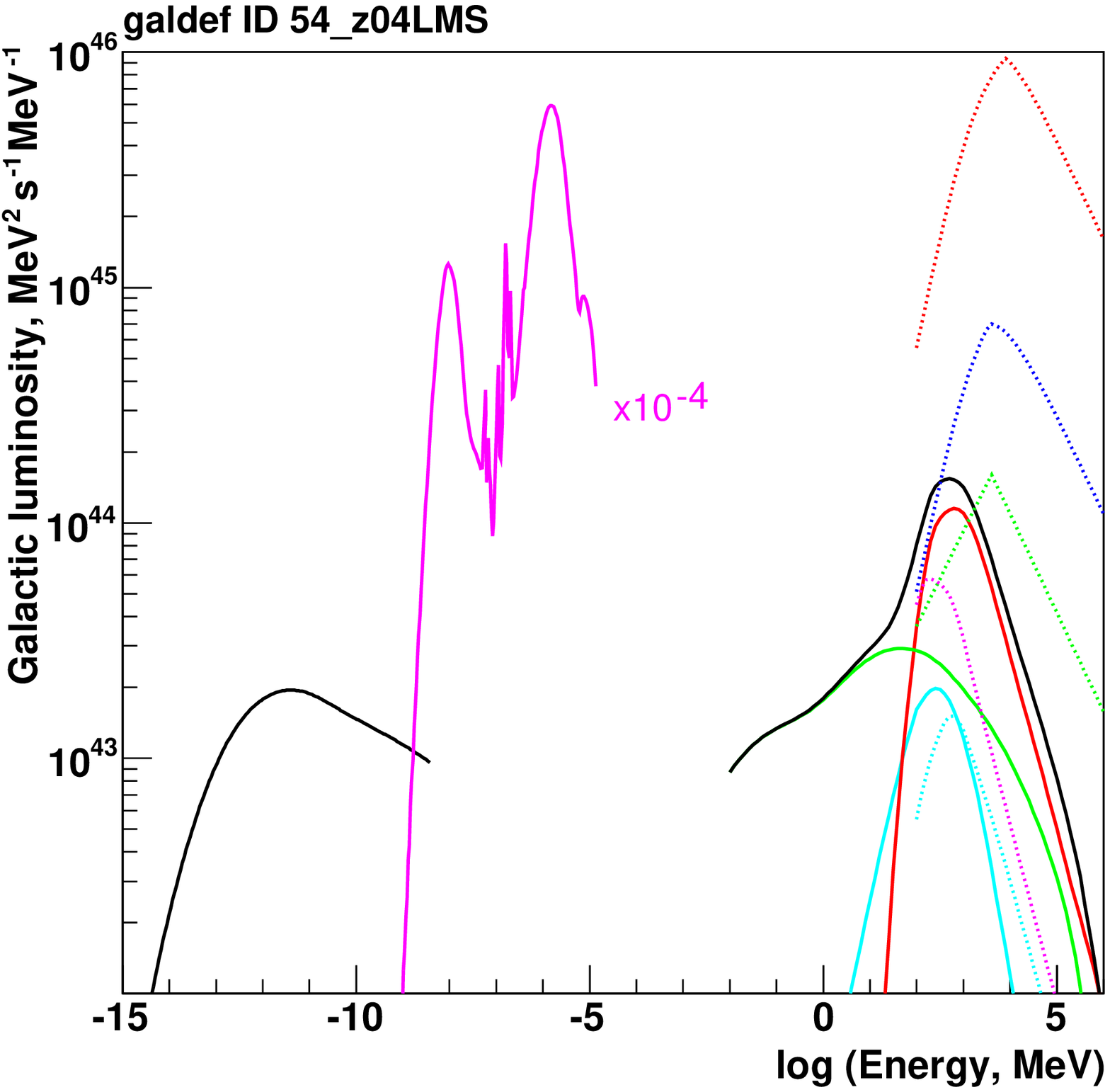} %\vspace{0.25in}
\includegraphics[width=3.75in]{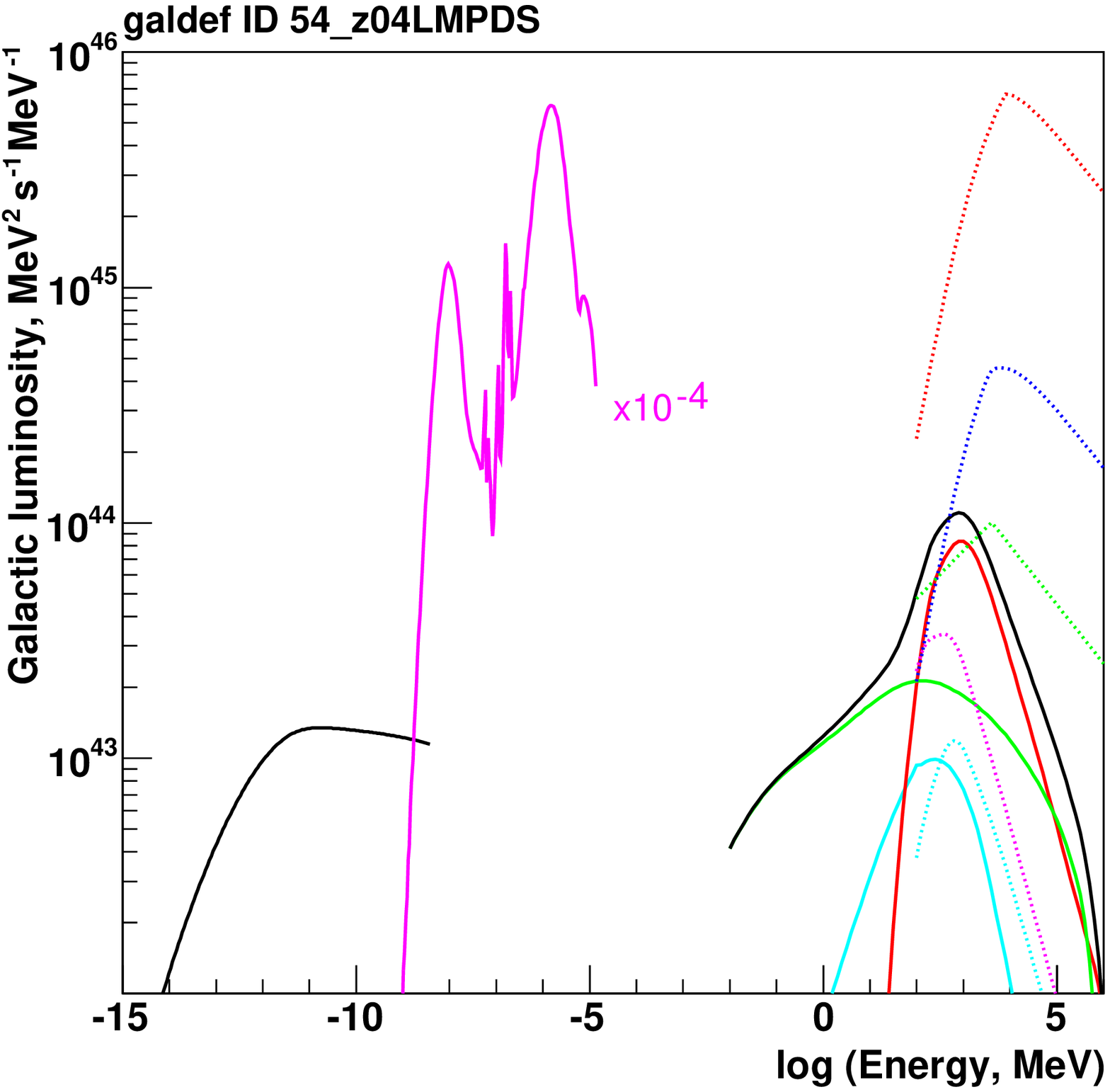}  %\vspace{0.25in}
}
\caption{Global CR-induced luminosity spectra of the MW 
for DR propagation model (left panel) 
and PD propagation model (right panel) with $z _h = 4$ kpc. 
Line styles: ISRF, including optical and 
infrared {\it scaled by factor $10^{-4}$} (magenta solid) and 
components for model 2 --  
Cosmic rays (dotted lines), 
protons (red), helium (blue), primary electrons (green), 
secondary electrons (cyan), secondary positrons (magenta); 
CR-induced diffuse emissions (solid lines), 
IC (green), bremsstrahlung (cyan), $\pi^0$-decay (red), 
synchrotron (black, left side of figure), total (black, right side of figure).}
\label{fig1}
\end{figure}

\begin{figure}
\centerline{\includegraphics[width=3.5in]{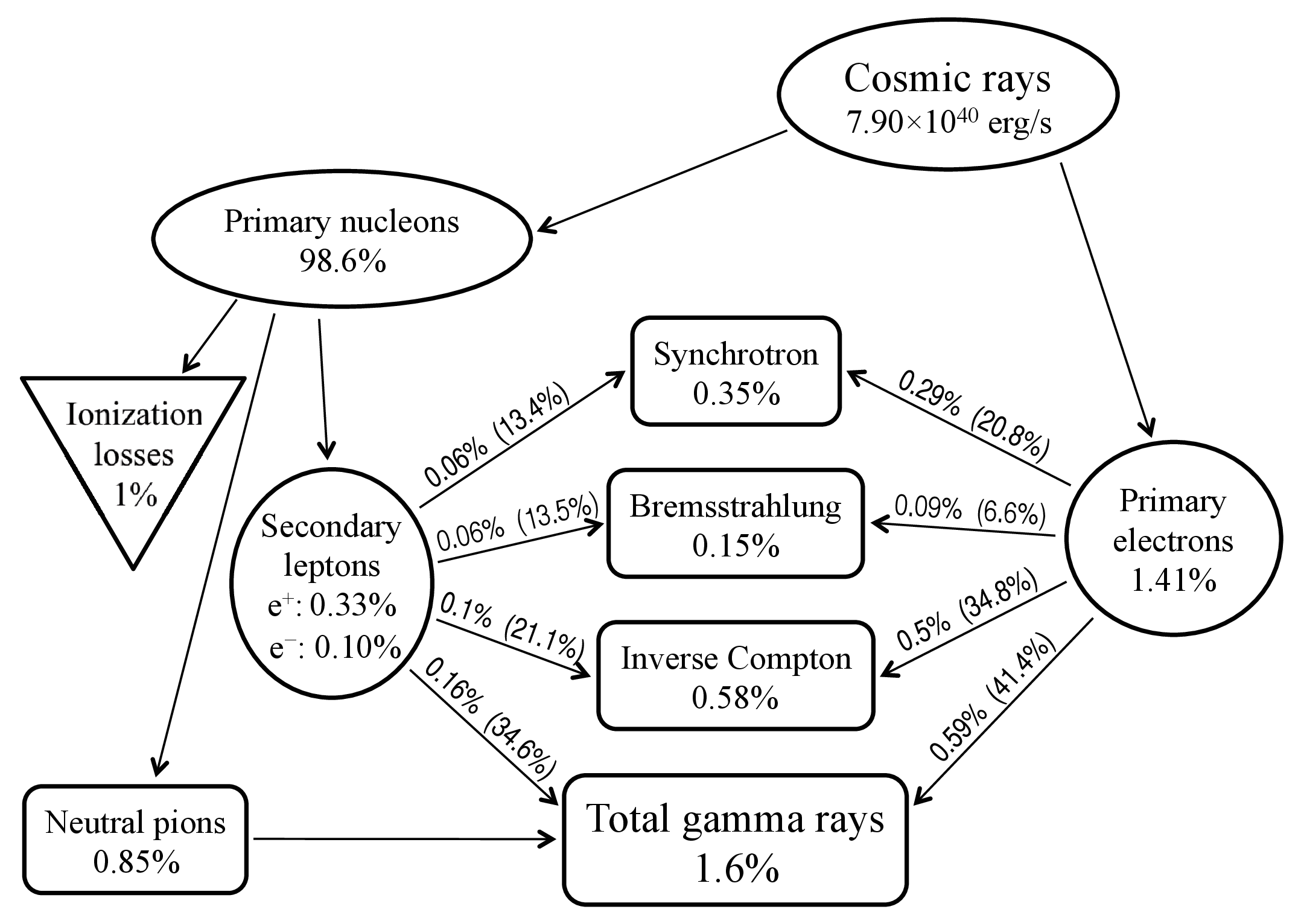}}%\vspace{0.25in}
\caption{The luminosity budget of the MW for DR propagation model 
with $z _h = 4$ kpc.
The percentage figures
are shown with respect to the total injected 
luminosity in CRs, $7.9\times10^{40}$ erg s$^{-1}$. 
The percentages in brackets show the values relative to the luminosity of 
their respective lepton populations 
(primary electrons, secondary electrons/positrons).}
\label{fig2}
\end{figure}

\begin{figure}
\centerline{
\includegraphics[width=3.75in]{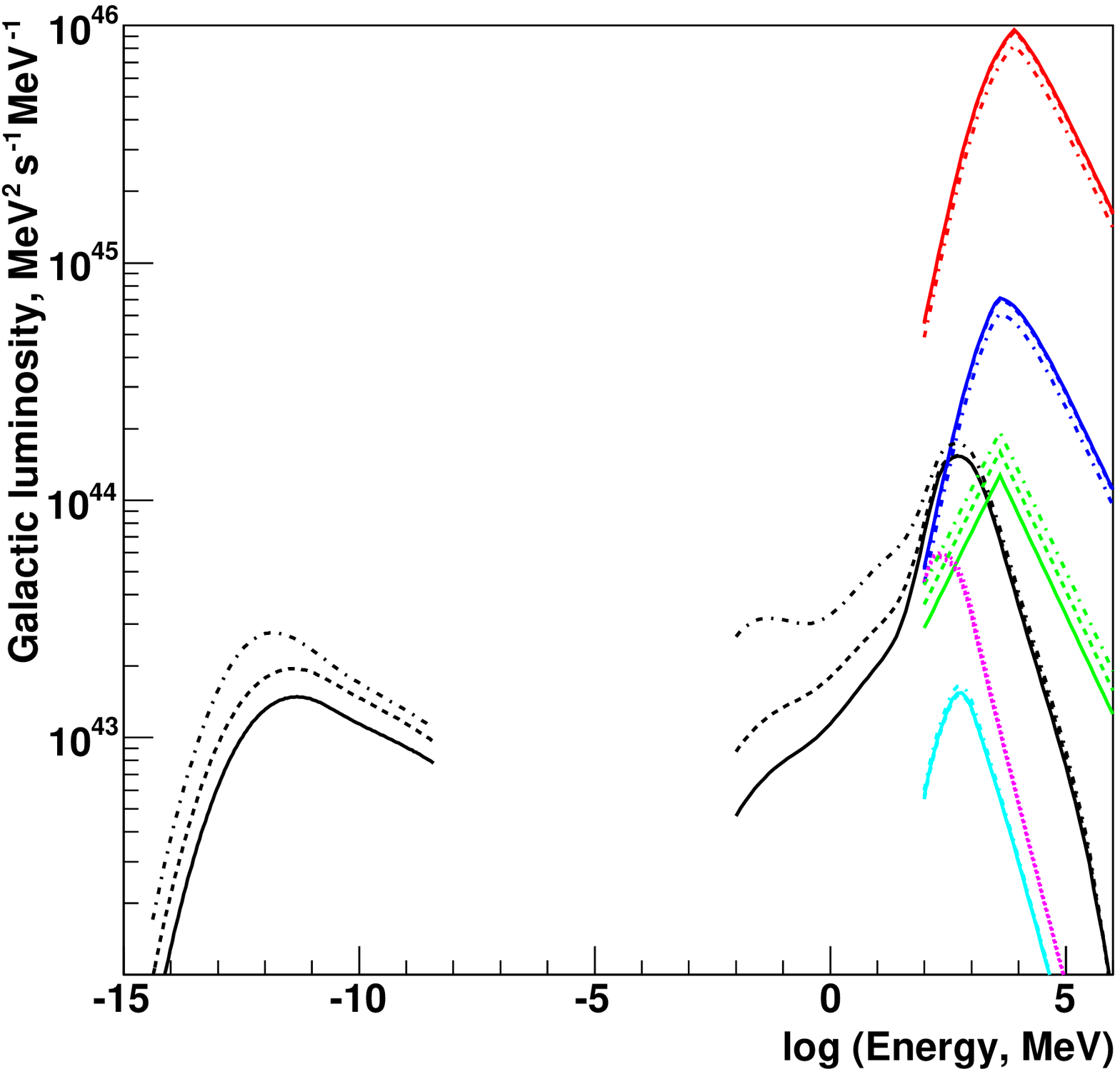} %\vspace{0.25in}
\includegraphics[width=3.75in]{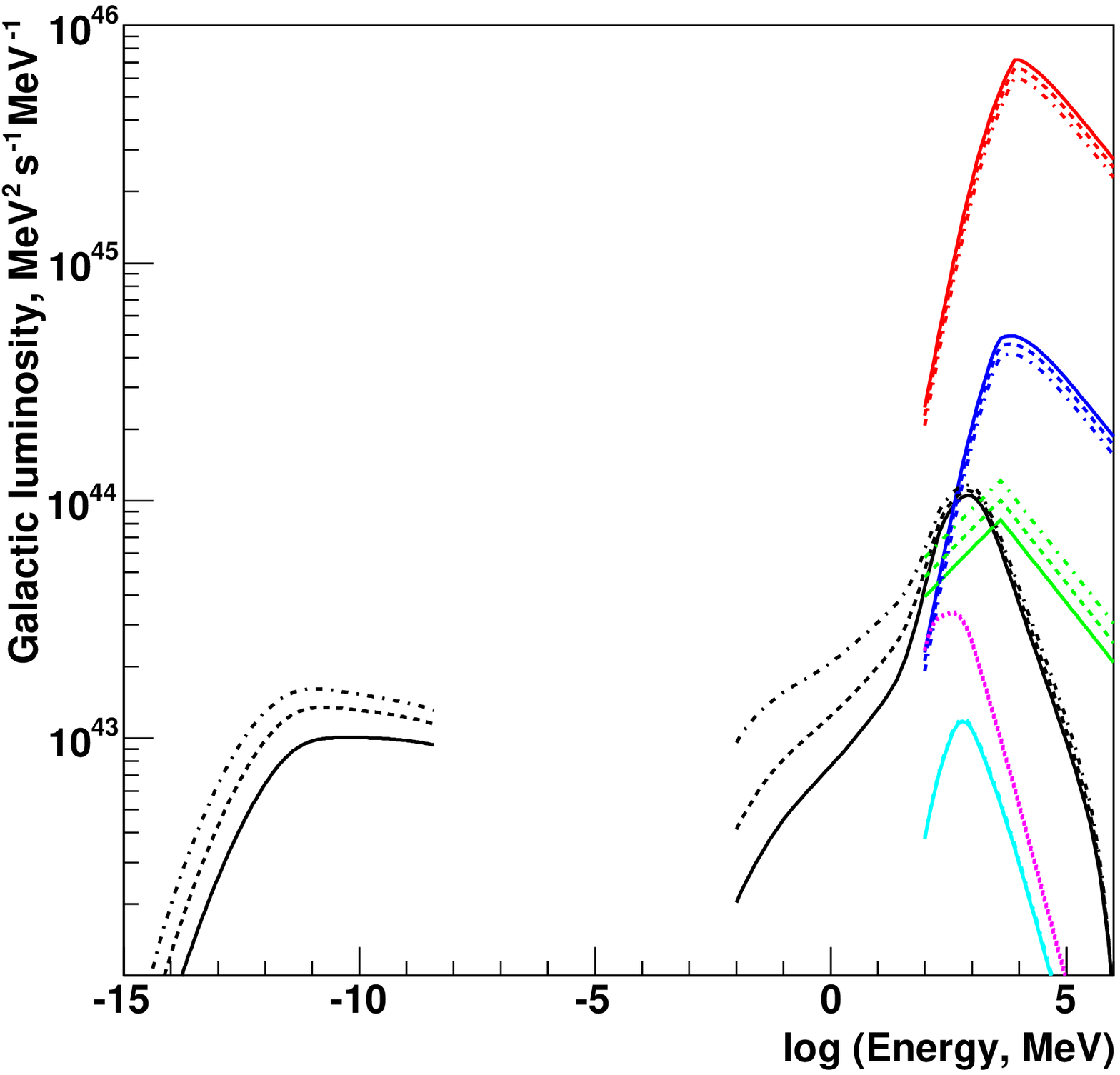}  %\vspace{0.25in}
}
\caption{Global CR-induced luminosity spectra of the MW for range
of halo sizes for DR propagation model (left panel) and 
PD propagation model (right panel).
Line styles: $z_h = 2$ kpc (solid), $z_h = 4$ kpc 
(dotted), and $z_h = 10$ kpc (dot-dashed). Components as in Fig.~\ref{fig1}.}
\label{fig3}
\end{figure}

Figure \ref{fig1} shows the broad-band luminosity 
spectrum of the Galaxy, including the input luminosity 
for CRs for a 4 kpc halo for a DR and PD model, 
respectively\footnote{Spectra for all models and components will be 
made available in numerical form  via the German Virtual Observatory, 
http://www.g-vo.org.}.
Fig.~\ref{fig2} illustrates the detailed energy budget for the DR model with
4 kpc halo size.
Figure~\ref{fig3} shows the broad-band luminosities for DR and PD 
models for the 3 halo sizes, and 
Table~\ref{table:2} summarises the spectrally-integrated luminosity for the 
various components for each of our models.
The peak injected CR luminosities differ by a factor $\sim 2$ for 
different propagation modes, but the total injected CR luminosities are 
close, only differing at the $\sim$few precent level.
The injected CR spectra differ between DR and PD models 
to compensate the momentum dependence of the diffusion coefficient in 
each propagation model so that the local CR spectra are reproduced.
Independent of propagation mode, the relative decrease in the injected 
CR proton and helium luminosities is $\sim10$\% 
for halo sizes 2 to 10 kpc.
For smaller halo sizes, the CRs escape quicker requiring more injected power 
to maintain the local CR spectrum.
In addition, for larger halo sizes CR sources located at further distances
can contribute to the local spectrum, which is our normalisation condition, 
hence less power is required.
In contrast to nuclei, the injected 
primary CR electron luminosity {\it increases} 
with $z_h$, 
reflecting the increased input power of these particles 
required to counter the energy losses from the 
larger confinement region and escape time.

The CR nuclei luminosities can be directly compared with the well-known 
approximate estimate from CR ``grammage'' as described 
by, e.g., \cite{Dogiel2002}, which shows these can be estimated 
from the observed secondary/primary ratios.
The values in Table~\ref{table:2} are within a 
factor 2 of the $\sim$$10^{41}$ erg s$^{-1}$ derived in that paper, 
providing a consistency check on our more detailed modelling.
Taking the CR luminosities and 
adopting a supernova rate of ($1.9\pm1.1)$/century 
\citep{Diehl2006} we obtain a CR energy input per SNR of 
$(0.3-1)\times 10^{50}$ erg, 
comparable to standard estimates of a 5--10\% efficiency for CR 
acceleration in SNRs.

For the DR model, 
the injected CR lepton luminosity is close to the total of synchrotron, IC, 
and bremsstrahlung luminosities for the 10 kpc halo (Model 3), showing that 
the Galaxy is approximately a lepton calorimeter ($\sim$$79$\% efficiency) 
in this case, with the bulk of the energy lost via IC emission (only about
one third is lost via synchrotron emission).
Smaller halos still constitute fair calorimeters but are less efficient:
$\sim$$51$\% for the case of a 2 kpc halo with $\sim$$17$\% of the energy 
loss due to synchrotron radiation, and $\sim$$59$\% for the case of a 4 kpc
halo with $\sim$$19$\% of the energy loss due to synchrotron radiation.
For PD models the calorimetric efficiencies are decreased, 
but the same trend of higher efficiency with increasing halo size is evident.

Our calculations show that calorimetry holds for CR leptons in MW-like galaxies
independent of propagation mode 
provided the size of the CR confinement volume is large enough to 
allow for sufficient cooling {\it and} if \gray{} production is included.
A large fraction of the IC luminosity is at energies below 
100 MeV, extending to hard X-rays, and includes a significant 
secondary electron/positron component \citep{Strong2004a,Porter2008}.
The contribution to the synchrotron emission from secondary 
electrons and positrons is $\sim10-20\%$ for our Galaxy-like models, 
being largest for small halo sizes.
For DR models, the secondary lepton emission is 20\%, 16\%, and 16\% for
2, 4, and 10 kpc, respectively.
For PD models, the secondary lepton emission is 12\%, 10\%, and 10\% for 
2, 4, and 10 kpc respectively. 

Using the standard definition of the 
FIR/radio relation, $q = \log_{10}(S_{\rm FIR}/3.75\times 10^{12} {\rm Hz}) - \log_{10}(S_{1.4 {\rm GHz}})$, where $S_{\rm FIR}$ (W m$^{-2}$) 
is the FIR flux 
from $42.5$--$122.5$ $\mu$m and $S_{1.4\ \rm GHz}$ (W m$^{-2}$ Hz$^{-1}$) 
is the radio flux
at 1.4 GHz \citep{Helou1985}, 
we calculate this ratio for our luminosity models (see Table~\ref{table:2}
under ``radio-FIR relation''). 
The range of values $2.26-2.69$ is 
consistent with that for the correlation over normal and 
starburst galaxies: $2.34 \pm 0.26$ \citep{Yun2001} 
and near to the value for M33 (2.50) given in that paper.
Note, here we have used the FIR luminosity from
the ISRF model described earlier, $1.9 \times 10^{43}$ 
erg s$^{-1}$ (see Table~\ref{table:2}). 
%$1.94\times 10^{36}$ W (convert to erg s$^{-1}$). 
However, the FIR luminosity of the Galaxy is uncertain by at least a factor 
2 and this will affect $q$. 
Using COBE/DIRBE data, \citet{Sodroski1997} give a total IR luminosity
$4.1 \times 10^{43}$ erg s$^{-1}$ and IR luminosity $>40$ $\mu$m, 
IR$_{> 40 \mu{\rm m}}$, 
of 
$2.8 \times 10^{43}$ erg s$^{-1}$.
This yields a FIR luminosity 
of $1.4 \times 10^{43}$ erg s$^{-1}$ 
from FIR/IR$_{> 40 \mu{\rm m}} = 0.51$ for the ISRF used in this paper. 
Meanwhile, \cite{Paladini2007} give a much higher value for 
IR$_{>60 \mu{\rm m}}$, $1.1 \times 10^{44}$ erg $^{-1}$, yielding 
a FIR luminosity $6.6 \times 10^{43}$ erg s$^{-1}$
(Paladini, private communication).
This higher value for the FIR luminosity would increase all $q$ values in 
Table~\ref{table:2} by 0.53.

The luminosity spectra for our models contains significant
IC \gray{} emission, the contribution at high energies depending
on details of the propagation model.
Above $\sim$$10$ GeV the contribution by IC is non-negligible compared to 
$\pi^0$-decay.
Estimates of the isotropic background using models for the diffuse \gray{} 
emission of normal star-forming galaxies invoking only $\pi^0$-decay as 
the dominant mode for high-energy production, e.g., \citet{Ando2009}, 
and ignoring details of the propagation will 
underestimate the contribution per galaxy by these classes of objects.

The present estimates are intended to be illustrative, using a particular
set of models; the level of uncertainty is suggested by the range covered
by the models which are far from exhaustive. 
In any case the general conclusions from this work are robust.

%%%%%%%%%%%%%%%
\acknowledgments
%%%%%%%%%%%%%%%
A.W.S. and T.A.P. would like to thank the organisers of 
the {\it Infrared Emission, 
Interstellar Medium, and Star Formation} Workshop 
(http://www.mpia-hd.mpg.de/IR10/) and the participants for the many 
stimulating talks and discussions.
We would also like to thank George Helou and Heinz V\"{o}lk for useful 
comments.
T.~A.~P. acknowledges support from NASA Grant No.~NNX10AE78G.
I.~V.~M. acknowledges support from NASA Grant No.~NNX09AC15G.

\clearpage

%%%%%%%%%%%%%%%%%%%%%%%%%%%%%%%%%%%%%%%%%%%%%%%%%%%%%%%%%%%%%%%%%%%%%%%%%%%%%%%
             %%%%%%%%%%%%%%%  T A B L E S  %%%%%%%%%%%%%%%%%
%%%%%%%%%%%%%%%%%%%%%%%%%%%%%%%%%%%%%%%%%%%%%%%%%%%%%%%%%%%%%%%%%%%%%%%%%%%%%%%

\begin{deluxetable}{lccccccc} 
\tabletypesize{\footnotesize}
\tablecaption{\label{table:1} GALPROP model parameters}
\tablecolumns{8}
\tablewidth{0pt}
\tablehead{
& \multicolumn{3}{c}{Diffusive Reacceleration} && \multicolumn{3}{c}{Plain Diffusion} \\
\cline{2-4} \cline{6-8} 
& Model 1 & Model 2 & Model 3 && Model 1 & Model 2 & Model 3\\ 
Model parameter/GALDEF ID & z02LMS & z04LMS & z10LMS && z02LMPDS & z04LMPDS & z10LMPDS}
\startdata
Halo height, kpc   & 2 & 4 & 10 && 2 & 4 & 10\\
Galaxy radius, kpc & 20 & 20 & 20 && 20 & 20 & 20 \\
Diffusion coefficient\tablenotemark{a} $D_0$ & 2.9  & 5.8  & 10.0 && 1.8  & 3.4  & 6.0\\
Diffusion coefficient\tablenotemark{a} $\delta$ & 0.33 & 0.33 & 0.33 && 0.5 & 0.5 & 0.5 \\
Reacceleration $v_A$, km s$^{-1}$ & 30  & 30 & 30 && \nodata & \nodata & \nodata \\
CR sources distribution &  \multicolumn{7}{c}{Pulsars \citep{Lorimer2004}\tablenotemark{b}}\\
Magnetic field strength\tablenotemark{c} & \multicolumn{7}{c}{$B = 7 e^{-(R-R_0)/R_B - z/z_B}$ $\mu$G}\smallskip\\
\multicolumn{4}{l}{\sc Injection spectrum (nuclei)} \\ 
\quad Index below break & 1.98 & 1.98 & 1.98 && 1.80 & 1.80 & 1.80 \\
\quad Index above break & 2.42 & 2.42 & 2.42 && 2.25 & 2.25 & 2.25 \\
\quad Break energy, GeV & 9 & 9 & 9 && 9 & 9 & 9 \\
\quad Normalisation energy, GeV & 100 & 100 & 100 && 100 & 100 & 100 \\
\quad Proton normalisation intensity\tablenotemark{d,e} & 5 & 5 & 5 && 5 & 5 & 5 \smallskip\\
\multicolumn{4}{l}{\sc Injection spectrum (primary electrons) }\\ 
\quad Index below break & 1.60 & 1.60 & 1.60 && 1.80 & 1.80 & 1.80 \\
\quad Index above break & 2.42 & 2.42 & 2.42 && 2.25 & 2.25 & 2.25 \\
\quad Break energy, GeV & 4 & 4 & 4 && 4 & 4 & 4\\
\quad Normalisation energy, GeV & 34.5 & 34.5 & 34.5 && 34.5 & 34.5 & 34.5 \\
\quad Normalisation intensity\tablenotemark{d} & 0.32 & 0.32 & 0.32 && 0.32 & 0.32 & 0.32 \\
\enddata
\tablenotetext{a}{$D_{xx}= 10^{28}\beta D_0 (\rho/\rho_0)^\delta$ cm$^2$ s$^{-1}$, $\rho_0=4$ GV, $\beta=v/c$, constant below $\rho_0$ for plain diffusion model.}
\tablenotetext{b}{Modified to have the the value at $R = 10$ kpc across the range $R = 10-15$ 
kpc and to be zero beyond 15 kpc.}
\tablenotetext{c}{$R_0 = 8.5$ kpc, $R_B = 50$ kpc, $z_B = 3$ kpc.}
\tablenotetext{d}{$10^{-9}$ cm$^{-2}$ s$^{-1}$ sr$^{-1}$ MeV$^{-1}$ (nucleon$^{-1}$ for nuclei).}
\tablenotetext{e}{Normalisation energy 100 GeV for protons corresponds to
a rigidity $\sim 100$ GV.
Helium normalisation is 0.068 relative to protons at the {\it same rigidity}.}
\end{deluxetable}

\begin{deluxetable}{lccccccc}
\tabletypesize{\scriptsize}
\tablecaption{\label{table:2} Luminosity of the Galaxy for various processes, $10^{38}$ erg s$^{-1}$.}
\tablecolumns{8}
\tablewidth{0pt}
\tablehead{
%& \multicolumn{6}{c}{Luminosity, $10^{38}$ erg s$^{-1}$} \\ 
& \multicolumn{3}{c}{Diffusive Reacceleration} && \multicolumn{3}{c}{Plain Diffusion} \\
\cline{2-4} \cline{6-8} 
Component & Model 1 & Model 2 & Model 3 && Model 1 & Model 2 & Model 3
}
\startdata
\sc Cosmic rays (0.1--100 GeV): & 805 & 790 & 698 && 780 & 723 & 660 \\
\quad Protons                   & 737 & 724 & 633 && 718 & 662 & 601 \\    % units of 1e38
\quad Helium                    & 56 & 55 & 48 && 52.4 & 48.3 & 43.9  \\%\smallskip\\ 
\quad Leptons                   & 12.2 & 14.5 & 16.9 && 10.04 & 12.91 & 15.15\\
\qquad \it Primary   e$^-$      & \quad \it 8.8 & \quad \it 11.1 & \quad \it 13.4 && \quad \it 8.65 & \quad \it 10.5 & \quad \it 12.7 \\
\qquad \it Secondary e$^-$      & \quad \it 0.78 & \quad \it 0.77 & \quad \it 0.83 && \quad \it 0.63 & \quad \it 0.64 & \quad \it 0.65 \\
\qquad \it Secondary e$^+$      & \quad \it 2.6 & \quad \it 2.6 & \quad \it 2.7  && \quad \it 1.76 & \quad \it 1.77 & \quad \it 1.80 \smallskip \\

\sc \grays\ (0.01--100 MeV):    & 2.32 & 3.34 & 6.22 && 1.47 & 2.20 & 3.50 \\
\quad $\pi^0$-decay             & 0.24 & 0.23 & 0.23 && 0.14 & 0.13 & 0.13 \\

\quad Inverse Compton           & 1.80 & 2.81 & 5.63 && 1.09 & 1.79 & 3.04 \\
\qquad \it Primary   e$^-$      & \quad \it 1.31 & \quad \it 2.20 & \quad \it 4.41 && \quad \it 0.91 & \quad \it 1.54 & \quad \it 2.67 \\
\qquad \it Secondary e$^\pm$    & \quad \it 0.49 & \quad \it 0.61 & \quad \it 1.22 && \quad \it 0.18 & \quad \it 0.25 & \quad \it 0.37 \\

\quad Bremsstrahlung            & 0.27 & 0.30 & 0.36 && 0.24 & 0.28 & 0.33 \\
\qquad \it Primary   e$^-$      & \quad \it 0.11 & \quad \it 0.15 & \quad \it 0.19 && \quad \it 0.15 & \quad \it 0.19 & \quad \it 0.24 \\
\qquad \it Secondary e$^\pm$    & \quad \it 0.16 & \quad \it 0.15 & \quad \it 0.17 && \quad \it 0.09 & \quad \it 0.09 & \quad \it 0.09 \smallskip \\

\sc \grays\ (0.1--100 GeV):    & 8.86 & 9.12 & 10.3 && 6.72 & 7.18 & 7.79 \\
\quad $\pi^0$-decay            & 6.75 & 6.46 & 6.59 && 4.99 & 4.90 & 4.79 \\

\quad Inverse Compton          & 1.25 & 1.76 & 2.59 && 1.27 & 1.77 & 2.43 \\
\qquad \it Primary   e$^-$     & \quad \it 1.15 & \quad \it 1.66 & \quad \it 2.43 && \quad \it 1.20 & \quad \it 1.68 & \quad \it 2.33 \\
\qquad \it Secondary e$^\pm$   & \quad \it 0.10 & \quad \it 0.10 & \quad \it 0.16 && \quad \it 0.07 & \quad \it 0.09 & \quad \it 0.10 \\

\quad Bremsstrahlung           & 0.87 & 0.88 & 1.08 && 0.46 & 0.51 & 0.57 \\
\qquad \it Primary   e$^-$     & \quad \it 0.51 & \quad \it 0.58 & \quad \it 0.74 && \quad \it 0.33 & \quad \it 0.39 & \quad \it 0.45 \\
\qquad \it Secondary e$^\pm$   & \quad \it 0.36 & \quad \it 0.30 & \quad \it 0.34 && \quad \it 0.13 & \quad \it 0.12 & \quad \it 0.12 \smallskip\\

\sc Radio (0.001--100 GHz):    & 2.07 & 2.76 & 3.72 && 1.48 & 2.03 & 2.50 \\ %synchrotron
\quad Primary e$^-$            & 1.65 & 2.31 & 3.10 && 1.30 & 1.81 & 2.26 \\
\quad Secondary e$^\pm$        & 0.42 & 0.45 & 0.62 && 0.18 & 0.22 & 0.24 \smallskip\\

\sc Conversion efficiencies:\\
\quad \grays/CR leptons\tablenotemark{a} & 0.34 & 0.40 & 0.57 && 0.28 & 0.34 & 0.42  \\
\quad Synchrotron/CR leptons\tablenotemark{b} & 0.17 & 0.19 & 0.22 && 0.13 & 0.16 & 0.17 \\
\quad Lepton calorimetric efficiency\tablenotemark{c} & 0.51 & 0.59 & 0.79 && 0.41 & 0.49 & 0.59 \smallskip\\

\sc Optical--FIR (0.3--3000 THz): & \multicolumn{7}{c}{$24.8\times 10^{43}$ erg s$^{-1}$} \\
\quad Optical\tablenotemark{d}  &  \multicolumn{7}{c}{$19.6\times 10^{43}$  erg s$^{-1}$}\\
\quad Total IR\tablenotemark{e}  &  \multicolumn{7}{c}{$5.2\times 10^{43}$  erg s$^{-1}$ }  \\
\quad FIR\tablenotemark{f} &  \multicolumn{7}{c}{$1.9\times 10^{43}$ erg s$^{-1}$ }  \smallskip \\

\sc Radio-FIR relation: \\
\quad Synchrotron (1.4 GHz)\tablenotemark{g}& 1.68 & 2.18 & 2.85 && 1.06 & 1.49 & 1.82\\
\quad $q_{\rm FIR}$ & 2.49 & 2.38 & 2.26 && 2.69 & 2.54 & 2.45\\
%\sc Synchrotron (1.4 GHz)\tablenotemark{g}& 1.68 & 2.18 & 2.85 && 1.06 & 1.49 & 1.82 \smallskip\\
%\sc Radio-FIR relation $q_{\rm FIR}$ & 2.49 & 2.38 & 2.26 && 2.69 & 2.54 & 2.45\\
\enddata

\tablenotetext{a}{(IC + bremsstrahlung)/(total leptons).}
\tablenotetext{b}{(Synchrotron)/(total leptons).}
\tablenotetext{c}{(IC + bremsstrahlung + synchrotron)/(total leptons).}
\tablenotetext{d}{0.1 -- 8 $\mu$m; $38-3000$ THz.}
\tablenotetext{e}{8 -- 1000 $\mu$m; $0.3-38$ THz.}
\tablenotetext{f}{42.5 -- 122.5 $\mu$m; $2.45-7.06$ THz.}
\tablenotetext{g}{$10^{28}$ erg s$^{-1}$ Hz$^{-1}$.}
\end{deluxetable}

\clearpage

%%%%%%%%%%%%%%%%%%%%%%%%%%%%%%%%%%%%%%%%%%%%%%%%%%%%%%%%%%%%%%%%%%%%%%%%%%%%%%%
             %%%%%%%%%%%%%%%  F I G U R E S  %%%%%%%%%%%%%%%%%
%%%%%%%%%%%%%%%%%%%%%%%%%%%%%%%%%%%%%%%%%%%%%%%%%%%%%%%%%%%%%%%%%%%%%%%%%%%%%%%

\end{document}